\documentclass{article}

\usepackage[utf8]{inputenc}
\usepackage{amsmath,amssymb,dsfont,hyperref}
\usepackage{cite,braket}
\usepackage{titlesec}
\usepackage[dvipsnames]{xcolor} 
\usepackage{cleveref}
\usepackage{slashed}
\usepackage{simpler-wick}
\usepackage{graphicx}
\usepackage{tikz-feynman}\tikzfeynmanset{compat=1.1.0}

\allowdisplaybreaks

\def \be {\begin{equation}}
\def \ee {\end{equation}}
\def \ba {\begin{eqnarray}}
\def \ea {\end{eqnarray}}

\def \tr {{\rm tr}_{\rm spin}}

\newcommand{\LJM}{{\mathcal{L}}}
\newcommand{\LJi}{{L}}

\usepackage{color}

\newcommand{\sub}[1]{\begin{subequations} #1 \end{subequations}}

\title{\LARGE\bf\boldmath GTMDs, orbital angular momentum, and pretzelosity}

\author{Brean Maynard, Peter Schweitzer \\
  \footnotesize \it Department of Physics, University of Connecticut, Storrs, CT 06269-3046, U.S.A.}

\date{\footnotesize April 2026}

\begin{document}

\maketitle

\begin{abstract}
The leading Generalized Transverse Momentum Dependent parton distributions (GTMDs) are 
studied in the bag model. The model description is shown to be theoretically consistent. 
The orbital angular momentum is studied in terms of the GTMD $F_{1,4}^q$ and Ji sum rule. 
Analytical proofs of the associated sum rules are given. A deeper relationship between 
orbital angular momentum and the pretzelosity TMD is established in this model. \\
\end{abstract}

%====== SECTION 1: INTRODUCTION ============================================
\section{Introduction}
\label{Sec-1:intro}

Generalized transverse momentum dependent parton distributions (GTMDs) emerged as an overarching concept
\cite{Meissner:2008ay,Meissner:2009ww} unifying the description of hadrons in terms of parton distribution 
functions (PDFs), transverse momentum dependent PDFs (TMDs) \cite{Boussarie:2023izj}, 
and generalized parton distribution functions (GPDs) \cite{Belitsky:2005qn}. GTMDs contain Wigner 
functions \cite{Ji:2003ak,Belitsky:2003nz} as a limiting case, providing a key to learn about the motion 
and spatial distribution of partons in the transverse plane. Certain GTMDs describe completely novel 
properties not associated with TMDs or GPDs. One such interesting property is parton orbital angular 
momentum \cite{Lorce:2011kd}. A pedagogical exposition of the interrelations between
PDFs, TMDs, GPDs, and GTMDs can be found in \cite{Lorce:2025aqp}.

The theory of the description of high energy processes in QCD in terms of PDFs, TMDs and GPDs has reached an advanced stage 
\cite{Collins:1989gx,Muller:1994ses,Ji:1996ek,Radyushkin:1996nd,Radyushkin:1996ru,Ji:1996nm,Collins:1996fb,Collins:1998be,Collins:2011zzd,Echevarria:2016scs},
though their phenomenological understanding differs from case to case, see e.g.\ \cite{Guo:2025muf,Boer:2025ixc,Liuti:2024zkc,Bacchetta:2024qre,Aslan:2024nqg,Cuic:2023mki,Moos:2023yfa,Dutrieux:2021nlz,Hou:2019efy,Moutarde:2019tqa}. 
In contrast to this, the situation for GTMDs is at a still pioneering stage with the first processes sensitive
to gluon or quark GTMDs proposed in 
Refs.~\cite{Ji:2016jgn,Hatta:2016aoc,Bhattacharya:2017bvs,Bhattacharya:2018lgm,Kovchegov:2024wjs,Bhattacharya:2026qnd}.
Our current knowledge of their nonperturbative properties is very limited. This makes GTMDs compelling objects for model studies.

Models can provide valuable lessons about hadron structure, in some instances 
crucially deepening our knowledge, e.g., see the review on the role of models 
for the understanding of TMDs in QCD in Sec.~7 of Ref.~\cite{Boussarie:2023izj}. 
In other~instances, one finds model-dependent but nevertheless interesting
insights. One example for this is the relation of orbital angular momentum to the 
pretzelosity TMD \cite{Ma:1998ar,Avakian:2008dz,She:2009jq} supported in a class 
of quark models 
\cite{She:2009jq,Avakian:2010br,Avakian:2010nz,Efremov:2010cy,Lorce:2011kd-plus}
which have a certain symmetry in common \cite{Lorce:2011kn}.
Although some models \cite{Liu:2014zla,Chakrabarti:2016yuw} and QCD do not support it,
this relation revealed a first connection of orbital angular momentum to TMDs and 
is one of the motivations for the present work.

In this work, we will study the leading order GTMDs of the nucleon 
% associated with the Dirac structure $\Gamma=\gamma^+$ 
known as $F_{1,1}^q$, $F_{1,2}^q$, $F_{1,3}^q$, $F_{1,4}^q$ in
the bag model \cite{Chodos:1974je,Chodos:1974pn,Johnson:1975zp}. 
This is a quark model, i.e.\ a model with no explicit gauge degrees of freedom,
a simplification allowing one to omit Wilson lines. In such models, time-reversal
odd TMDs vanish \cite{Pobylitsa:2002fr} and GTMDs are real-valued. The bag model 
has a long track record of often serving for the first exploratory investigations 
of hadronic properties including 
structure functions \cite{Jaffe:1974nj,PhysRevC.28.1256}, 
transversity PDFs \cite{Jaffe:1991kp}, 
GPDs \cite{Ji:1997gm,Scopetta:2005fg,Tezgin:2024tfh},
TMDs \cite{Avakian:2008dz,Avakian:2010br,Courtoy:2008vi}, 
double parton distributions \cite{Chang:2012nw, Rinaldi:2013vpa},
and orbital angular momentum \cite{Avakian:2010br,Burkardt:2010he,Courtoy:2016des}.

After introducing GTMDs and their properties in Sec.~\ref{Sec-2} and the bag model in Sec.~\ref{Sec-3}, 
we will derive the expressions for $F_{1,1}^q$, $F_{1,2}^q$, $F_{1,3}^q$, $F_{1,4}^q$ in the bag model
and demonstrate their theoretical consistency in Sec.~\ref{Sec-4}. Hereby we will work in the large-$N_c$ 
limit which is necessary to consistently satisfy the Ji sum rule \cite{Ji:1997gm,Neubelt:2019sou}.
We will discuss the orbital angular momentum related to $F_{1,4}^q$ in Sec.~\ref{Sec-5}
and the one obtained from the Ji sum rule in Sec.~\ref{Sec-6},
and prove the latter analytically in Sec.~\ref{Sec-7}. Finally, in Sec.~\ref{Sec-8} we will establish
a deeper relationship between orbital angular momentum and pretzelosity through the framework of GTMDs,
before presenting in Sec.~\ref{Sec-9} the concluding remarks.

We remark that the GTMD $F_{1,4}^q$ was studied previously in the bag model using the Peierls-Yoccoz 
projection in \cite{Courtoy:2016des}. Studies of nucleon GTMDs in other model can be found in 
Refs.~\cite{Lorce:2011dv,Lorce:2011ni,Courtoy:2013oaa,Kanazawa:2014nha,Mukherjee:2015aja,
Liu:2015eqa,Zhou:2016rnt,Hagiwara:2016kam,More:2017zqq,Chakrabarti:2019wjx,
Gurjar:2021dyv,Maji:2022tog,Sharma:2023tre,Chakrabarti:2024hwx,Luo:2020yqj,Tan:2024dmz}. 

\newpage
\section{Definition and properties of leading GTMDs}
\label{Sec-2}

In quark models with no gluon degrees of freedom, the fully unintegrated 
quark-quark correlator is defined as \cite{Meissner:2009ww}
\begin{equation}
    W_{s s^{\prime}}^{[\Gamma]q}\left(P,k,\Delta\right)
    =\frac12\int \frac{d^4z}{(2 \pi)^4} e^{i k \cdot z} 
    \bra{ p^{\prime},s^{\prime}}\bar{\psi}^q(-\tfrac{z}{2}) 
    \,\Gamma\, \psi^q(\tfrac{z}{2})\ket{ p,s }
    \label{Eq:correlator-0}
\end{equation}
where $P=\frac12(p'+p)$, $\Delta=p'-p$, and nucleon states are normalized covariantly as 
$\langle p',s^{\prime} | p,s \rangle =2p^0\,\delta^{(3)}(\vec{p}^{\,\prime}-\vec{p}^{\,})\delta_{s 's }$. We define the light-cone components of a generic 
four-vector $a^\mu$ as $a^\mu=(a^+,\vec{a}_T,a^-)$ with $a^\pm=\frac{1}{\sqrt{2}}(a^0\pm a^3)$. 
In this notation, the GTMDs related to $\Gamma=\gamma^+$ are defined as \cite{Meissner:2009ww}
\begin{align}
    {\cal M}^{[\gamma^{+}]q}_{s s '} \,
    & = \iint dk^+dk^-\,\delta(k^+-xP^+)
        W_{s s^{\prime}}^{[\gamma^{+}]q}(P,k,\Delta) \nonumber\\
    & = \frac{1}{2M} \bar{u}\left(p^{\prime}, s^{\prime}\right)
         \left[F_{1,1}^q
        +\frac{i \sigma^{i+} k_T^i}{P^{+}} F_{1,2}^q
        +\frac{i \sigma^{i+} \Delta_T^i}{P^{+}} F_{1,3}^q
        +\frac{i \sigma^{i j} k_T^i \Delta_T^j}{M^2} F_{1,4}^q\right] u(p, s ),
        \label{Eq:correlator-1}
\end{align}
where the nucleon spinors are normalized as 
$\bar{u}(p, s^{\prime})u(p, s )=2M\delta_{s 's }$ and $M$ denotes the
nucleon mass.
The GTMDs are in general complex valued and depend on the 5 variables 
$x, \; \xi, \; \vec{k}_T^2, \; \vec{k}_T\cdot\vec{\Delta}_T,\;\vec{\Delta}_T^2$ 
with $\xi$ and $t$ defined as \cite{Meissner:2009ww}
\begin{equation}
    \xi = -\,\frac12\;\frac{\Delta^+}{P^+} = \frac{p^+-p^{\prime+}}{p^++p^{\prime+}}\,, \quad \quad
    t = \Delta^2 = -\,\frac{\xi^2(4M^2+\vec{\Delta}_T^{\,2})}{1-\xi^2} - \vec{\Delta}_T^{\,2}\,.
\end{equation}
The variable $t$ will be needed below for GPDs. 

GTMDs posses multiple properties of particular interest. 
Hermiticity of the correlator implies the property \cite{Meissner:2009ww}
\begin{equation}
    F_{1,j}^{q\,*}(x, \xi, \vec{k}_T^2,  \vec{k}_T\cdot\vec{\Delta}_T,\vec{\Delta}_T^2)
    = \pm F_{1,j}^q(x, -\xi, \vec{k}_T^2,  -\vec{k}_T\cdot\vec{\Delta}_T,\vec{\Delta}_T^2)
    \quad \text{with} \quad
    \begin{cases}
       + & \text{for} \ j = 1,\;3,\;4,\\ 
       - & \text{for} \ j = 2.
    \end{cases}
    \label{Hermiticity}
\end{equation}
% with a plus sign for $j = 1,\;3,\;4$ and a minus sign for $j=2$. 
In the forward limit $\Delta^\mu\to 0$, the variables
$\xi, \; \vec{k}_T\cdot\vec{\Delta}_T,\;\vec{\Delta}_T^2$ become zero.
In this limit, the GTMD $F_{1,1}^q$ is related to the unpolarized TMD 
\cite{Bacchetta:2006tn} as follows \cite{Meissner:2009ww}
\begin{equation}
    F_{1,1}^q(x, 0,\vec{k}_T^2,0, 0) = f_1^q(x,\vec{k}_T^2). \label{TMDLimit}
\end{equation}
Integrating out transverse quark momenta, the GTMDs  $F_{1,1}^q$, $F_{1,2}^q$, $F_{1,3}^q$
give rise to the unpolarized GPDs as follows
\sub{
\begin{align}
     H^q(x, \xi, t)  =
     & \int d^2 \vec{k}_T\left[F_{1,1}
     +2 \xi^2\left(\frac{\vec{k}_T \cdot \vec{\Delta}_T}{\vec{\Delta}_T^2}
     F_{1,2}^q+F_{1,3}^q\right)\right]\label{HGPD}\\
     E^q(x, \xi, t) =
     & \int d^2 \vec{k}_T\left[-F_{1,1}^q+2\left(1-\xi^2\right)
     \left(\frac{\vec{k}_T \cdot \vec{\Delta}_T}{\vec{\Delta}_T^2}
     F_{1,2}^q+F_{1,3}^q\right)\right]\label{EGPD}.
\end{align}}
The GTMD $F_{1,4}^q$ is of particular interest because upon integration over $x$ and 
transverse quark momenta with the weight $\vec{k}_T^2/M^2$ it gives access to the contribution 
of quark orbital momentum to the nucleon spin according to \cite{Lorce:2011kd}
\begin{equation}\label{OAM}
        \LJM_z^q
        = -\int\limits_{-1}^1 dx \int d^2 k_T \;\frac{\vec{k}_T^{\,2}}{M^2} 
        \,F_{1,4}^q\left(x, 0, \vec{k}_T^{\,2}, 0,0\right) \,,
\end{equation}
which is often referred to as the canonical or Jaffe-Manohar orbital angular 
momentum because, through a gauge-invariant extension, it is related to the 
Jaffe-Manohar nucleon spin decomposition \cite{Jaffe:1989jz}.

By choosing the Dirac matrices $\Gamma = \gamma^+\gamma_5$ or $i\sigma^{j+}\gamma_5$
yields respectively the leading GTMDs $G_{1,j}$ with $1\le j \le 4$ or $H_{1,j}$ 
with $1\le j \le 8$. Choosing Dirac matrices other than that yields subleading 
GTMDs \cite{Meissner:2009ww}. In this work, we will focus on the GTMDs related to 
$\Gamma=\gamma^+$ and feature one of the GTMDs related to $\Gamma=i\sigma^{j+}\gamma_5$. 
A study of all GTMDs in the bag model will be presented elsewhere \cite{new}.

\newpage
\section{The bag model}
\label{Sec-3}

In the bag model, $N_c=3$ non-interacting quarks are confined in a color singlet  state 
within a spherical cavity (``bag'') of radius $R$ by means of appropriate boundary conditions 
\cite{Chodos:1974je,Chodos:1974pn,Johnson:1975zp}. The single-quark ground state wave function 
for a massless quark in momentum space is given by 
\begin{equation}\label{Eq:wf-mom}
        \Psi_s(t, \vec{k})=e^{-i \varepsilon_0 t} \,A 
        \binom{t_0(k)\chi_s}{t_1(k) \,\vec{k} \cdot \vec{\sigma}\,\chi_s},
\end{equation}
where $\varepsilon_0=\omega_0/R$ is the ground-state energy, 
$\omega_0\approx 2.04$ is the lowest positive solution of the 
transcendental equation $\omega = (1-\omega)\tan{\omega}$, and 
$\sigma^i$ are the Pauli matrices with $\chi_s$ denoting the 
two-component Pauli spinors.  
The functions $t_{\ell}$ for $\ell=0,1$ are defined in terms of 
spherical Bessel functions $j_{\ell}$ as follows
\begin{equation}
    t_{\ell}(k)=\frac{\int_0^1 du\; u^2j_{\ell}(ukR)j_{\ell}(u\omega_0)}{k^\ell}\,.
\end{equation}
Notice that we define $t_1(k)$ with a factor of $1/k$ which differs 
from what is often used in literature such that in our work $t_0(k)$ is 
dimensionless while $t_1(k)$ has dimension 1/mass. This definition
will simplify the notation later on. The wave function is normalized as
\be  
    \int d^3k\;\Psi_{s'}^{\dagger}(\vec{k})\Psi_s^{ }(\vec{k})=
    A^2 \int d^3k\bigl(t_0^2+\vec{k}^2t_1^2\bigr)\delta_{ss'}
    = \delta_{ss'}\label{eq:Normalization}
\ee
with the normalization constant 
$A =\frac{1}{2\pi}\bigl[\bigl(\omega_0R^3\bigr)/\bigl((\omega_0-1)j_0^2(\omega_0)\bigr)\bigr]^{\!1/2}$.

The bag radius is fixed in terms of the nucleon mass $M$ via the 
relation $M=\frac43N_{c\,}\epsilon_0$. The calculations are conveniently 
carried out for ``flavorless'' quarks with the flavor dependence 
assigned based on the SU(4) spin-flavor symmetry by means of the 
spin-flavor factors \cite{PhysRevD.30.238}, with $N_q$ for spin-independent and 
$P_q$ for spin-dependent nucleon matrix elements, given for the proton by 
(with $u$ and $d$ being interchanged for the case of the neutron)
\begin{equation}\label{Eq:spin-flavor-factors}
    N_u=\frac{N_c+1}{2},\quad N_d=\frac{N_c-1}{2}, \quad 
    P_u=\frac{N_c+5}{6},\quad P_d=\frac{-N_c+1}{6}\,.
\end{equation}
For more detailed introductions to the model, we refer to Refs.~\cite{Chodos:1974je,Chodos:1974pn,Johnson:1975zp}.

\section{Calculation of GTMDs in the model}
\label{Sec-4}

In the bag model, it is convenient to use the Breit frame where
$P^\mu=(P^0,0,0,0)$ and $\Delta^\mu = (0,\vec{\Delta})$ and 
$t=-\vec{\Delta}^2$.
In order to ensure the validity of the Ji sum rule in the bag model, 
we will work in the large-$N_c$ limit \cite{Ji:1997gm,Neubelt:2019sou}
where $M={\cal O}(N_c)$ and $P^0=M+{\cal O}(N_c^{-1})$ and we will neglect
${\cal O}(N_c^{-1})$ corrections. The behavior of the variables is 
\be 
    k^i_T       = {\cal O}(N_c^0)   \, , \quad
    \Delta^i_T  = {\cal O}(N_c^0)   \, , \quad 
    x           = {\cal O}(N_c^{-1})\, , \quad 
    \xi = -\Delta^3/(2M) = {\cal O}(N_c^{-1})\, , \quad
    t = {\cal O}(N_c^0)\,.
\ee
The expressions in Eq.~(\ref{Eq:correlator-1}) can be understood as
$2\times2$ matrices in nucleon helicity indices. In order to evaluate
these expressions, it is convenient to take traces with the unit and 
Pauli matrices. Evaluating in this way in the large $N_c$ limit the nucleon
spinor expressions on the right-hand side of Eq.~(\ref{Eq:correlator-1}) for 
``flavorless quarks'' yields the results
\sub{\label{Eq:correlator-GTMDs-large-Nc}
\begin{align}
   \tfrac12\,\tr\left[{\cal M}^{[\,\gamma^{+}]}\;\mathds{1}\;\right]
    & = \phantom{-\,}F_{1,1} \, , 
    \\
   \tfrac12\,\tr\left[{\cal M}^{[\gamma^{+}]}\;\sigma_x\right]
    & = \phantom{-\,}\frac{ik_y}{M}\,F_{1,2}+\frac{i\Delta_y}{M}\,F_{1,3} \, , 
    \\
   \tfrac12\,\tr\left[{\cal M}^{[\gamma^{+}]}\;\sigma_y\right]
    & =-\,\frac{ik_x}{M}\,F_{1,2}\,-\,\frac{i\Delta_x}{M}\,F_{1,3} \, , 
    \\
   \tfrac12\,\tr\left[{\cal M}^{[\gamma^{+}]}\;\sigma_z\right]
    & =\phantom{-\,}\frac{i\bigl(k_x\Delta_y-k_y\Delta_x\bigr)}{M^2} \,F_{1,4} \, ,
\end{align}}
where we write for more clarity $\vec{k}=(k_x,k_y,k_z)$ and similarly for 
$\vec{\Delta}$ and neglect terms suppressed in large-$N_c$ limit.
Taking into account the quark flavor factors (\ref{Eq:spin-flavor-factors}),
the large-$N_c$ scaling of GTMDs obtained from this expansion reads 
\begin{equation}\label{Eq:GTMD-large-Nc-counting}
    F_{1,1}^q = N^q \mathcal{O}(N_c), \qquad 
    F_{1,2}^q = P^q \mathcal{O}(N_c^2), \qquad 
    F_{1,3}^q = P^q \mathcal{O}(N_c^2), \qquad 
    F_{1,4}^q = P^q \mathcal{O}(N_c^3). 
\end{equation}
Note that different flavor combinations have different $N_c$-orders. 
In unpolarized case, the isoscalar flavor combination $N^u+N^d=N_c$ is 
leading while the isovector flavor combination $N^u-N^d=1$ is subleading. 
In polarized case it is vice versa with $P^u+P^d=1$ and $P^u-P^d=\frac{N_c+2}3$. 
Our choice to apply the large-$N_c$ expansion in 
Eq.~(\ref{Eq:correlator-GTMDs-large-Nc}) before projecting on definite 
flavors simplifies the model evaluation and corresponds to the scheme 
used in \cite{Neubelt:2019sou}.\footnote{\label{footnote1} In an alternative 
    large-$N_c$ scheme, one could include the SU(4) spin-flavor 
    in Eq.~(\ref{Eq:correlator-GTMDs-large-Nc}) before expanding the 
    spinor expressions. Then one would proceed with separate model 
    calculations for the $(u\pm d)$ cases. In the strict large-$N_c$ 
    limit, both schemes yield the same results.}

The correlator on the left-hand side of Eq.~(\ref{Eq:correlator-1}) 
is given in the bag model (for flavorless quarks) in the large-$N_c$ limit
by the expression 
\begin{equation}
    {\cal M}^{[\gamma^{+}]}_{s s '} = M\;\bar{\phi}_{s '}(\vec{k}+\tfrac12\vec{\Delta}) (\gamma^0+\gamma^3)
    \phi_s (\vec{k}-\tfrac12\vec{\Delta}) \Bigr|_{k_z=xM-\epsilon_0} \,.
    \label{Eq:correlator-gamma+}
\end{equation}
where we indicate that $k_z=xM-\varepsilon_0$ is fixed.
Evaluating this expression in the model yields the results
\sub{\label{Eq:correlator-model}
\begin{align}
    \tfrac12\,\tr\left[\,{\cal M}^{[\gamma^{+}]}\;\mathds{1}\;\right]
    & = MA^2\Bigl[ 
        t_0^- t_0^+ 
      + (k_z-\tfrac12\Delta_z)t_1^-t_0^+ 
      + (k_z+\tfrac12\Delta_z)t_0^-t_1^+
      + (\vec{k}^{\,2}-\tfrac14\,\vec{\Delta}^2)t_1^-t_1^+\Bigr]_{k_z=xM-\epsilon_0}
    \,,\\
    \tfrac12\,\tr\left[{\cal M}^{[\gamma^{+}]}\,\sigma_x\right]
    & = MA^2\Bigl[  
      i\,k_y(t_0^-t_1^+ - t_1^-t_0^+ - \Delta_z t_1^-t_1^+)
      +i\,\Delta_y(\tfrac12t_0^-t_1^+ + \tfrac12t_1^-t_0^+ + k_zt_1^-t_1^+)\Bigr]_{k_z=xM-\epsilon_0}
    \,,\\
    \tfrac12\,\tr\left[{\cal M}^{[\gamma^{+}]}\,\sigma_y\right]
    & = MA^2\Bigl[ 
      i\,k_x(t_1^-t_0^+ - t_0^-t_1^+ + \Delta_zt_1^-t_1^+)
    - i\,\Delta_x(\tfrac12t_0^-t_1^+ + \tfrac12t_1^-t_0^+ +  k_zt_1^-t_1^+) \Bigr]_{k_z=xM-\epsilon_0}
    \,,\\
    \tfrac12\,\tr\left[{\cal M}^{[\gamma^{+}]}\,\sigma_z\right]
    & = MA^2\Bigl[-\,i\,(k_x\Delta_y-k_y\Delta_x)t_1^+t_1^-\Bigr]_{k_z=xM-\epsilon_0}
\end{align}}
where we introduced the notation
\begin{equation}
    t_\ell^\pm = t_\ell\left(|\vec{k}\pm\tfrac12\vec{\Delta}|\right) \, .
\end{equation}
Comparing the coefficients in 
Eqs.~(\ref{Eq:correlator-GTMDs-large-Nc},~\ref{Eq:correlator-model}) 
we obtain the following results for the GTMDs
\sub{\label{Eqs:F1j-model}
\begin{align}
    F_{1,1}^q & = N^q A^2 M\;\Bigl[ 
        t_0^- t_0^+  + k_z(t_0^-t_1^+ + t_1^-t_0^+)
      + \tfrac12\Delta_z(t_0^-t_1^+ - t_1^-t_0^+)    
      + (\vec{k}^{\,2}-\tfrac14\,\vec{\Delta}^2)t_1^-t_1^+
      \Bigr]_{k_z=xM-\epsilon_0}\,,
      \label{Eq:F11-model}\\
    F_{1,2}^q & = P^q A^2M^2\Bigl[\;t_0^-t_1^+ - \; t_1^-t_0^+ - \Delta_z t_1^-t_1^+
      \Bigr]_{k_z=xM-\epsilon_0}\,,
      \label{Eq:F12-model}\\
    F_{1,3}^q & = P^q A^2M^2\Bigl[\tfrac12t_0^-t_1^+ + \tfrac12t_1^-t_0^+ + k_zt_1^-t_1^+
      \Bigr]_{k_z=xM-\epsilon_0}\,,
    \label{Eq:F13-model}\\
    F_{1,4}^q & = P^q A^2 M^3\Bigl[-t_1^+t_1^- \Bigr]_{k_z=xM-\epsilon_0} \, .
    \label{Eq:F14-model}
 \end{align}}
In order to demonstrate the theoretical consistency of the model expressions, 
we first note that 
\be
    |\vec{k}\pm\tfrac12\vec{\Delta}| = 
    \sqrt{\vec{k}_T^{\,2}+\tfrac14\vec{\Delta}_T^{\,2}\pm\vec{\Delta}_T\cdot\vec{k}_T
    +(k_z\pm\tfrac12\Delta_z)^2}\, , \quad
    k_z = xM - \epsilon_0 , \quad
    \Delta_z = - 2\xi M\, ,
\ee
which shows that the model GTMDs are explicit functions of the variables 
$F_{1,j}=F_{1,j}(x, \, \xi, \, \vec{k}_T^{\,2}, \, \vec{k}_T\cdot\vec{\Delta}_T,\, \vec{\Delta}_T^{\,2})$.
We recall that $t_0^\pm$ is dimensionless while $t_1^\pm$ has dimension 1/mass.
Therefore, all GTMDs have the correct dimension 1/mass$^2$ and one obtains (in the specific cases)
dimensionless GPDs upon integration over transverse momenta. Next we note that replacing $\xi\to (-\xi)$ 
and $\vec{\Delta}_T\to (-\vec{\Delta}_T)$ has the effect of $t_\ell^\pm \to t_\ell^\mp$.
This observation immediately shows us that the model results in 
Eqs.~(\ref{Eq:F11-model}-\ref{Eq:F14-model}) comply with the hermiticity 
property according to Eq.~(\ref{Hermiticity}).

Setting $\xi\to 0$ and $\vec{\Delta}_T\to0$ in the model expression for 
$F_{1,1}^q$ in Eq.~(\ref{Eq:F11-model}) implies $t_\ell^\pm \to t_\ell$ and
we obtain
\begin{equation}
    F_{1,1}^q(x,0,\vec{k}_T^2,0,0)
    = N^q A^2 M\;\Bigl[ t_0^2 + 2k_z t_0t_1 + \vec{k}^{\,2}t_1^2\Bigr]_{k_z=xM-\epsilon_0}   
    = f_1^q(x,\vec{k}^2_T) \, ,\label{TMDresults}
\end{equation}
where in the last step we recovered the expression for the unpolarized TMD 
$f_1^q(x,\vec{k}_T^{\,2})$ from Ref.~\cite{Avakian:2010br}.
In order to check the relation to GPDs we first note that in the large-$N_c$ 
limit the Eqs.~(\ref{HGPD},~\ref{EGPD}) simplify as
\sub{\label{Eq:H+EGPD-2}\begin{align}
     H^q(x, \xi, t) =
     & \int d^2 \vec{k}_T\;F_{1,1}^q\label{Eq:HGPD-2}\\
     E^q(x, \xi, t) =
     & \int d^2 \vec{k}_T\left[-F_{1,1}^q+2\,\frac{\vec{k}_T \cdot \vec{\Delta}_T}{\vec{\Delta}_T^2} F_{1,2}^q+2F_{1,3}^q\right]\label{Eq:EGPD-2}. 
\end{align}}
Inserting in Eq.~(\ref{Eq:H+EGPD-2}) the expressions for the $F_{1,j}$ 
from Eqs.~(\ref{Eqs:F1j-model}), we obtain results for the GPDs which are in agreement 
with Ref.~\cite{Ji:1997gm} if one considers the large-$N_c$ limit, see also 
\cite{Tezgin:2020vhb}.

\newpage
\section{Orbital angular momentum from GTMD \boldmath $F_{1,4}^q$}
\label{Sec-5}

It is customary to define an $x$-dependent orbital angular momentum distribution
function $\LJM_z^q(x)$ in terms of $F_{1,4}^q$ with $\xi\to0$ and $\Delta_T^i\to 0$ 
as follows
\ba
        \LJM_z^q(x) = - \int d^2 k_T \;\frac{\vec{k}_T^{\,2}}{M^2} \,
        F_{1,4}^q\left(x, 0, \vec{k}_T^{\,2}, 0,0\right) 
        \label{Eq:Lzq-Jaffe-Manohar} \,,
\ea
such that upon integration over $x$, one obtains $\LJM_z^q = \int dx\,\LJM_z^q(x)$ 
in Eq.~(\ref{OAM}). When evaluating $x$-integrals in the bag model,
it is convenient to proceed as follows ($R(\vec{k})$ below denotes a generic
function of $\vec{k}$ and $\vec{\Delta}$)
\be\label{Eq:x-integration-in-bag}
      M \int\limits_{-1}^1\! dx \int\! d^2k_T\bigr[R(\vec{k})\bigl]_{k_z=xM-\epsilon_0}
    = M \int\limits_{-1}^1\! dx \int\! d^3k \, R(\vec{k}) \delta(k_z+\epsilon_0-xM)
    = \int\! d^3k\,{R}(\vec{k}) \,.
\ee
The final result is obtained by substituting $x\to y = xM$ and using the large-$N_c$ 
limit to replace the integration interval $-M\le y \le M$ by the whole $y$-axis. 
In this way, using the bag model expression for
$F_{1,4}^q(z,0,\vec{k}_T^2,0,0)$ from Eq.~(\ref{Eq:F14-model}) and simplifying 
$\int d^3k\;k_i^{\,2}\,t_1^2 = \frac13\int d^3k\;\vec{k}^2\,t_1^2$ for $i=\{x,y,z\}$, 
we obtain
\be\label{Eq:OAM-Lzq}
    \LJM_z^q = P^q A^2 \int d^3 k \;
    \Bigl[\,\tfrac23\,\vec{k}^{\,2}\,t_1^2 \,\Bigr] \,.
\ee
This is the contribution to the nucleon spin originating from the canonical
quark orbital angular momentum.

To demonstrate the theoretical consistency of the model, we need to take 
into account the quark spin contribution $S^q$ to the nucleon spin which 
is defined in terms of the helicity PDF as
\be\label{Eq:Sq-helicity}
    S^q = \frac12 \int dx\,g_1^q(x)\,, \qquad 
    g_1^q(x) = \int d^2k_T\,g_1^q(x,k_T)\,
\ee
where we expressed the helicity PDF in terms of the helicity TMD $g_1^q(x,k_T)$,
which is possible in simple quark models though not in QCD \cite{Boussarie:2023izj}. 
The latter is given in the bag model by \cite{Avakian:2008dz,Avakian:2010br}
\be\label{Eq:g1-TMD}
    g_1^q(x,k_T) = P^q A^2 M\;\Bigl[ t_0^2 + 2k_z t_0t_1 
    - (\vec{k}_T^{\,2}-k_z^2)t_1^2\Bigr]_{k_z=xM-\epsilon_0} \,.
\ee
Evaluating $S_z^q$ in the bag model according to Eq.~(\ref{Eq:x-integration-in-bag}) 
where the term proportional to $k_z$ in Eq.~(\ref{Eq:g1-TMD}) drops out due to symmetry 
reasons, we obtain 
\be\label{Eq:Sq-2}
    S_z^q = \tfrac12 \int dx \int d^2k_T \;g_1^q(x,k_T) 
    = \tfrac12\;P^q A^2 \int d^3k\;\Bigl[ t_0^2 - \tfrac13\,\vec{k}^2\,t_1^2\Bigr] \,.
\ee
Adding up the results in Eqs.~(\ref{Eq:OAM-Lzq},~\ref{Eq:Sq-2}) and making use of 
the normalization in Eq.~(\ref{eq:Normalization}) yields for the total angular momentum 
contribution $J_z^q$ to the nucleon spin the result
\be\label{Eq:OAM-3}
        S_z^q+\LJM_z^q = \tfrac12\; P^q A^2 \int d^3 k \;
            \Bigl[\,t_0^2+ \vec{k}^2\,t_1^2 \,\Bigr] 
            =\tfrac12\,P^q\,.
\ee
If we sum over $u$- and $d$-quarks with the SU(4) spin-flavor 
factors in Eq.~(\ref{Eq:spin-flavor-factors}) we obtain 
$\sum_q (S_z^q+\LJM_z^q) = \frac12$ as expected, because in the bag model the 
nucleon spin is entirely due to the spin and orbital angular momentum of
quarks.

Our result for $2\LJM_z^u(x)$ is shown in Fig.~\ref{fig1}a and that
for $g_1^u(x)$ in Fig.~\ref{fig1}b. These results refer to a low hadronic scale 
estimated to be below $1\,\rm GeV^2$ \cite{Stratmann:1993aw}.
The results for $d$-flavor have opposite signs and are 4 times smaller 
according to the spin-flavor factors in Eq.~(\ref{Eq:spin-flavor-factors}). 
$F^q_{1,4}$ was studied previously in bag model without using large-$N_c$ 
limit in \cite{Courtoy:2016des}. Our results agree well with \cite{Courtoy:2016des}
indicating that $1/N_c$-corrections are moderate in the case of $F^q_{1,4}$.
In the bag model, the quark spin contributes $\sum_q 2S^q = 0.65$, i.e.\ $65\%$ 
to the nucleon spin, while the angular momentum contribution is 
$\sum_q 2\LJM_z^q = 0.35$, i.e.\ $35\%$. These are typical results in quark models
at low scales \cite{Myhrer:2007cf,Thomas:2008ga}. 

Before ending this section, we note that in QCD the $x$-integrals extend over  
$-1\le x \le 1$, with the understanding that negative $x$ describe antiquarks, 
e.g.\ $g_1^q(-x) = g_1^{\bar q}(x)$ in Eq.~(\ref{Eq:Sq-helicity}). The antiquark
distributions in the bag model are unphysical but must be included when evaluating 
sum rules \cite{Ji:1997gm}. Note also that the bag model gives rise to small but 
non-zero contributions outside the region $|x|<1$, which must also be included 
when evaluating sum rules \cite{Ji:1997gm}.

\newpage
\section{Orbital angular momentum from Ji sum rule}
\label{Sec-6}
A different (historically earlier) way to define quark orbital angular momentum 
is based on the Ji sum rule which states that the total angular momentum 
contributions $J_z^q$ of quarks can be accessed via GPDs as \cite{Ji:1996ek}
\be\label{Eq:Ji-sum-rule-0}
    J^q = \tfrac12\,\lim\limits_{t\to 0} 
    \int dx\,x\,\Bigl(H^q(x,\xi,t)+E^q(x,\xi,t)\Bigr)\,.
\ee
By generalizing the integrand in Eq.~(\ref{Eq:Ji-sum-rule-0}) to an
$x$-dependent distribution of total angular momentum and subtracting the 
spin contribution described by one half of the helicity PDF $g_1^q(x)$, 
it was proposed to define the orbital angular momentum distribution 
$\LJi_z^q(x)$ in the Ji decomposition as \cite{Hoodbhoy:1998yb},
see also the reviews \cite{Leader:2013jra,Liu:2015xha},
\be\label{Eq:Lzq-Ji}
    \LJi_z^q(x)  = \tfrac12\,x(H^q+E^q)(x,0,0)-\tfrac12\,g_1^q(x)\,.
\ee
In quark models (but not in QCD), both orbital angular momentum distributions 
have the same normalization \cite{Burkardt:2012sd},
\be
    \int dx\,\LJi_z^q(x) = \int dx\,\LJM_z^q(x) = \LJM_z^q \,,
\ee
but their $x$-dependence is in general different,
$\LJi_z^q(x) \neq \LJM_z^q(x)$.

\begin{figure}[b!]
  \begin{center}
    \includegraphics[height=6cm]{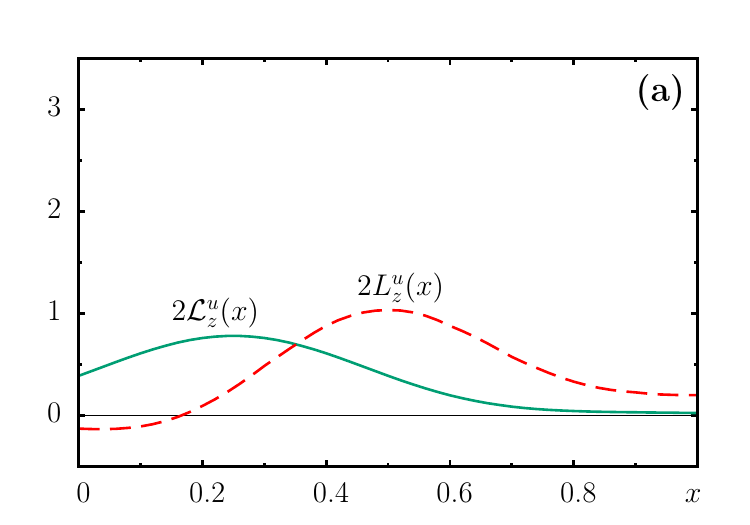} 
    \includegraphics[height=6cm]{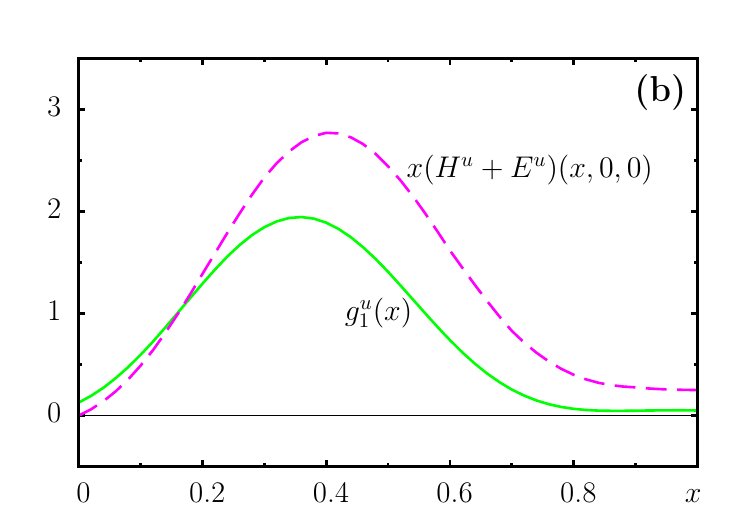}
  \end{center}
  \caption{\label{fig1}
  (a) Orbital angular momentum distributions in the bag model: 
  $2\LJM_z^u(x)$ obtained from GTMD $F_{1,4}^q$ 
  via Eq.~(\ref{Eq:Lzq-Jaffe-Manohar}) and
  $2\LJi_z^u(x)$ obtained from Ji decomposition
  via Eq.~(\ref{Eq:Lzq-Ji}).
  (b) For comparison: helicity PDF $g_1^u(x)$ and 
  twice the total angular momentum distribution $(H^u+E^u)(x,0,0)$.
  The results refer to a low scale below 1 GeV$^2$.} 
\end{figure}

To investigate this matter, we need the model expression for $H^q(x,\xi,t)+E^q(x,\xi,t)$
in the forward limit $\xi\to0$ and $t\to 0$. Taking this limit in Eq.~(\ref{Eq:H+EGPD-2}) 
yields
\sub{\label{Eq:GPD-forward-limit}
\ba H^q(x,0,0) &=& \phantom{-}f_1^q(x) \, 
    \label{Eq:GPD-H-forward-limit}\\
    E^q(x,0,0) &=& -f_1^q(x)+ P^qA^2M^2\int d^2k_T\,\biggl[
    \,\frac{\vec{k}_T^2}{k}\bigl(t_0t_1^\prime - t_0^\prime t_1\bigr) +
    2\biggl(t_0 t_1 + k_z t_1^2\biggr)\biggr]_{k_z=xM-\epsilon_0}. 
    \label{Eq:GPD-E-forward-limit}
\ea}
where $f_1^q(x)=\int d^2k_Tf_1^q(x,k_T)$ with $f_1^q(x,k_T)$ given by 
Eq.~(\ref{TMDresults}). The derivatives $t_\ell^\prime=dt_\ell(k)/dk$ in 
Eq.~(\ref{Eq:GPD-E-forward-limit}) come from applying l'H\^opital's rule to the 
term $\vec{k}\cdot\vec{\Delta}_T F_{1,2}/\vec{\Delta}_T^{2}$ in Eq.~(\ref{Eq:EGPD-2})
in the limit $\vec{\Delta}_T\to0$ (notice that in the model, $F^q_{1,2}$ is real and
therefore an odd function of $\Delta_T^i$ due to the property (\ref{Hermiticity}), 
which guarantees that the limit exists). 

From the expressions in Eqs.~(\ref{Eq:g1-TMD},~\ref{Eq:Lzq-Ji},~\ref{Eq:GPD-forward-limit}) 
we obtain the result for $2\LJi^u_z(x)$ shown in Fig.~\ref{fig1}a. For comparison, 
we display the forward limit $x(H^u+E^u)(x,0,0)$ in Fig.~\ref{fig1}b. The results for
$d$-quarks are again of opposite signs and four times smaller according to 
Eq.~(\ref{Eq:spin-flavor-factors}). In Fig.~\ref{fig1}a we see that also in the 
bag model $\LJi_z^q(x) \neq \LJM_z^q(x)$.

The Ji sum rule is satisfied in the bag model which was checked numerically 
in Refs.~\cite{Ji:1997gm,Neubelt:2019sou,Courtoy:2016des}. Given the importance
of this sum rule, we present an analytical proof in the next section which will 
show an important difference to the spin sum rule in Eq.~(\ref{Eq:OAM-3}).

\newpage
\section{Proof of Ji sum rule}
\label{Sec-7}

There is an important difference between the spin sum rule in Eq.~(\ref{Eq:OAM-3})
and the Ji sum rule. In order to illustrate this, we present here explicit analytical
proofs of the quark flavor, momentum and Ji sum rule which are given by
\ba
     \int dx f_1^q(x) &=& N^q, \label{Eq:flavor-sum-rule}\\
     \int dx \sum_q x\,f_1^q(x) &=& 1, \label{Eq:momentum-sum-rule}\\
     \int dx \; x(H^q+E^q)(x,0,0) &=& 2J^q. \label{Eq:Ji-sum-rule} 
\ea
To evaluate the quark flavor sum rule (\ref{Eq:flavor-sum-rule})
we proceed as in Eq.~(\ref{Eq:x-integration-in-bag}) whereby 
an odd term (under $k_z\to-k_z$) drops out, and we obtain
$\int dx f_1^q(x) = N^q A^2 \int d^3k\bigl(t_0^2+\vec{k}^2t_1^2\bigr)$
which proves the sum rule (\ref{Eq:flavor-sum-rule}) owing to the
wave function normalization in Eq.~(\ref{eq:Normalization}).
The proof of the spin sum rule in Eq.~(\ref{Eq:OAM-3}) is conceptually 
similar in that it can be traced back to the normalization of the quark 
wave function. In other words, in the bag model the proofs of these two 
sum rules rely on the normalization of the quark wave function in 
Eq.~(\ref{eq:Normalization}) and SU(4) spin-flavor factors. This alone is 
not sufficient to prove the sum rules (\ref{Eq:momentum-sum-rule}) 
and (\ref{Eq:Ji-sum-rule}).

In order to evaluate the momentum sum rule, we use the expression for 
the unpolarized TMD in Eq.~(\ref{TMDresults}) and carry out steps which
are analogous to those in Sec.~\ref{Sec-5}, which yields 
\be\label{Eq:proof-mom-sum-rule-1}
    \int dx\;x\,f_1^q(x) = \frac{N^q A^2}{M}\;\int d^3k
    \Bigl[\epsilon_0(t_0^2+\vec{k}^{\,2}t_1^2) + \tfrac23k^2 t_0t_1\Bigr]  . 
\ee
In order to proceed, we explore the equation of motion which is given 
in coordinate space by $(i\slashed{\partial}-m_q)\Psi_s(t,\vec{r})=0$ 
for $|\vec{r}^{\,}|<R$ where 
$\Psi_s(t,\vec{r})=\int d^3k\,e^{-i(\epsilon_0t-\vec{k}\cdot\vec{r})}\phi(\vec{k})$ 
is the Fourier transform of the flavorless wave function in Eq.~(\ref{Eq:wf-mom}) 
(we use the same notation for the function and its Fourier transform) and $m_q$ is the quark 
mass which we will eventually set to zero. With the understanding that $\Psi_s(t,\vec{r})=0$
for $|\vec{r}^{\,}|>R$ the following trivial identity
\be\label{Eq:identity-generic}
    \int d^3r\,F(t,\vec{r})\Gamma(i\slashed{\partial}-m_q)\Psi_s(t,\vec{r})=0
\ee
holds for any function $F(t,\vec{r})$ and any Dirac matrix $\Gamma$. Choosing
$F({t,}\vec{r})=\bar{\Psi}_{s'}(t,\vec{r})=
\int d^3k'\,e^{i(\epsilon_0t-\vec{k}'\!\cdot\vec{r})}\bar{\phi}(\vec{k}')$ 
and $\Gamma=\mathds{1}$ yields (so far we kept $m_q\neq0$ for didactic
purposes to indicate the Dirac equation, but now we set it to zero)
\ba
    \;\;\int\!\frac{d^3r}{(2\pi)^3}\,\bar{\Psi}_{s'}(t,\vec{r})(i\slashed{\partial}-m_q)\Psi_s(t,\vec{r})
    = \int\!\!d^3k\,\bar{\phi}_{s'}(\vec{k})(\epsilon_0\gamma^0-\vec{k}\cdot\vec{\gamma})\phi_s(\vec{k}) 
    = A^2\!\!\int\!\!d^3k \Bigl[\epsilon_0(t_0^2+\vec{k}^{\,2}t_1^2) - 2k^2 t_0t_1\Bigr]\delta_{ss'} = 0,
    \nonumber
\ea
which implies the identity 
$\int d^3k\,[2k^2 t_0t_1] = \epsilon_0\int d^3k (t_0^2+\vec{k}^{\,2}t_1^2)$.
Making use of this identity in Eq.~(\ref{Eq:proof-mom-sum-rule-1}) yields 
\be\label{Eq:proof-mom-sum-rule-2}
    \int dx\;x\,f_1^q(x) = N^q\,\frac{4\,\epsilon_0}{3M}\,A^2\!\!\int d^3k
    \Bigl[t_0^2+\vec{k}^{\,2}t_1^2\Bigr] = \frac{N^q}{N_c}
\ee
where in the last step we used Eq.~(\ref{eq:Normalization}) and $M=\frac43N_{c\,}\epsilon_0$. 
Summing over flavors yields the momentum sum rule (\ref{Eq:momentum-sum-rule}). 

Turning our attention to the Ji spin sum rule, we first note that  
$H^q(x,0,0)$ and $E^q(x,0,0)$ both contain a contribution from $f_1^q(x)$ but with 
opposite signs, see Eqs.~(\ref{Eq:GPD-H-forward-limit},~\ref{Eq:GPD-E-forward-limit}). 
Thus, the only piece in these GPDS that could have been literally related to the 
momentum sum rule cancels out exactly in the Ji spin sum rule which appears to be
a manifestly genuine off-forward effect in the sense that nothing known from the
forward case -- not even in a simple quark model -- can help one to prove this sum 
rule. However, we shall see that upon invoking an additional ingredient, namely the 
explicit consideration of off-forward kinematics, a proof of the Ji spin sum rule 
can be formulated analogously to the momentum sum rule. Note that both sum rules 
are ultimately rooted in the conservation of the energy momentum tensor, 
but in complimentary ways.

Starting with the model expressions for $H^q(x,0,0)$ and 
$E^q(x,0,0)$ in Eqs.~(\ref{Eq:GPD-forward-limit}) and using symmetry arguments yields 
\be\label{Eq:towards-Ji-1}
    \int dx\,x\bigl(H^q(x,0,0)+E^q(x,0,0)\bigr) = 
    P^q A^2\,\int d^3k\Bigl[\tfrac23\,\epsilon_0 k(t_0 t_1^\prime-t_0^\prime t_1)
        + 2\epsilon_0 t_0 t_1 + \tfrac23\,\vec{k}^2 t_1^2\Bigr].
\ee
Next we use Eq.~(\ref{Eq:identity-generic}) with 
$F({t,}\vec{r})=\bar{\Psi}_{s'}(t,\vec{r})\,e^{-i\vec{\Delta}\cdot\vec{r}}$ and 
$\Gamma = \gamma^3$. After evaluating the model expression and tracing with the 
unit matrix in nucleon spin indices (and setting $m_q=0$) we obtain the identity
\ba\label{Eq:towards-Ji-2}
    K(\vec{\Delta}^{\,})=A^2\,\int d^3k\Bigl[
    2k_z\bigl(-t_0^- t_0^+ 
    + \epsilon_0 t_0^+t_1^- 
    - \epsilon_0 t_0^-t_1^+
    + t_1^-t_1^+\bigr)    
    + \Delta_z \bigl(t_0^- t_0^+ 
    - \epsilon_0 t_0^+t_1^- 
    - \epsilon_0 t_0^-t_1^+
    + t_1^-t_1^+\bigr)   
    \Biggr] 
    = 0\,.
\ea
Taking the derivative of the above expression with respect to $\Delta_z$
and setting $\vec{\Delta}$ to zero yields the identity
\be\label{Eq:towards-Ji-3}
    \frac{\partial K(\vec{\Delta}^{\,})}{\partial\Delta_z}\biggl|_{\vec{\Delta}=0}
    = A^2\int d^3k\biggl[
    t_0^2 - 2\epsilon_0 t_0 t_1 + \tfrac13 k^2 t_1^2 
    + \tfrac23 \epsilon_0 k t_0^\prime t_1 
    - \tfrac23 \epsilon_0 k t_0 t_1^\prime
    \biggr] = 0\,.
\ee
Multiplying Eq.~(\ref{Eq:towards-Ji-3}) by $P^q$ and adding this ``zero''
to Eq.~(\ref{Eq:towards-Ji-1}) we obtain
\be\label{Eq:towards-Ji-4}
    \int dx\,x\bigl(H^q(x,0,0)+E^q(x,0,0)\bigr) = 
    P^q A^2\,\int d^3k\Bigl[t_0^2 + \vec{k}^2 t_1^2 \Bigr] = P^q\,,
\ee
where we made use of Eq.~(\ref{eq:Normalization}) in the final step. This 
proves the Ji sum rule in the bag model.

Note that in the proof of the {momentum sum rule (\ref{Eq:momentum-sum-rule})}, 
we had to make use of the bag model expression for the
nucleon mass $M=\frac43N_{c\,}\epsilon_0$. It is instructive to recall how this
result is obtained. Namely, it follows from minimizing the nucleon mass as a function 
of $R$, $M(R)=N_c \epsilon_0+\frac43\pi R^3B$ where $\epsilon_0=\omega_0/R$ and $B$ 
is the bag constant. The result $M=\frac43N_{c\,}\epsilon_0$ follows from setting 
$M'(R)=0$ and eliminating $B$. The condition $M'(R)=0$ is the virial theorem in the
bag model and is equivalent to the von Laue condition, which is ultimately connected to 
the conservation of the energy-momentum tensor in the bag model \cite{Neubelt:2019sou}
and in general theory \cite{Polyakov:2018zvc}.

This illustrates the important difference between the spin (\ref{Eq:OAM-3}) and
quark flavor (\ref{Eq:flavor-sum-rule}) sum rules on the one hand and the momentum 
(\ref{Eq:momentum-sum-rule}) and Ji (\ref{Eq:Ji-sum-rule}) sum rules on the other hand. 
The former merely require the normalization condition for the quark wave functions 
(\ref{eq:Normalization}), i.e.\ are based on the mere construction of quark flavor 
and spin quantum numbers within the SU(4) spin-flavor symmetry underlying this model.
The latter require in addition to that the use of the bag model equation of motion for 
the single quarks, $(i\slashed{\partial}-m_q)\Psi_s(t,\vec{r})=0$, {i.e.\ explicitly 
invokes the dynamics of the model. For the momentum sum rule, one further ingredient is 
needed, namely the determination of the nucleon mass, $M'(R)=0$, where all $N_c$ quarks 
collectively contribute along with the bag constant.}

Finally, we discuss the sum rule for the anomalous gravitomagnetic moment 
related to the energy-momentum tensor form factor $B(t)$ \cite{Ji:1996ek}. Based on the
results for the momentum sum rule (\ref{Eq:proof-mom-sum-rule-2}) and Ji sum rule
(\ref{Eq:towards-Ji-4}), we obtain
\be\label{Eq:gravitomangnetic-1}
    B^q(0) = \int dx\,x\,E^q(x,0,0) = P^q - \frac{N^q}{N_c}\,.
\ee
Numerically, we obtain the results $B^u(0) = \frac23$ and $B^d(0) = -\frac23$
for the proton and three colors. Summing over all flavors in the bag model, we 
find that the total anomalous gravitomagnetic moment of the nucleon vanishes, 
\be
    B(0) = \sum_q B^q(0) = 0\,,
\ee
which is expected and equivalent to the Ji sum rule \cite{Ji:1996ek}. 

At this point, it is instructive to comment on the large-$N_c$ limit, which we
used when deriving the results in (\ref{Eq:correlator-GTMDs-large-Nc}). 
Thereafter, we included in Eq.~in (\ref{Eqs:F1j-model}) the spin-flavor factors 
(\ref{Eq:spin-flavor-factors}), left $N_c$ general, and set it equal to its 
physical value when presenting results in Fig.~\ref{fig1} which constitutes 
our ``large-$N_c$ expansion scheme'' as described below 
Eq.~(\ref{Eq:GTMD-large-Nc-counting}).
One could be concerned if this is a consistent scheme. In fact, other schemes
could be applied as well, see footnote~\ref{footnote1}. A good test of the 
theoretical consistency of a model calculation is compliance with sum rules,
and we have explicitly demonstrated that our results correctly satisfy them.
Moreover, inserting the spin-flavor factors (\ref{Eq:spin-flavor-factors}) in
Eq.~(\ref{Eq:gravitomangnetic-1}) yields the $1/N_c$ expansion of the quark 
contributions to the anomalous gravitomagnetic moment of the proton
\be\label{Eq:gravitomangnetic-2}
    B^u(0) = -B^d(0) = \tfrac16\,N_c +  \tfrac13\,N_c^0 -  \tfrac12\,N_c^{-1} \,.
\ee
This shows that one may use our scheme or truncate the expansion after leading 
(or subleading) order. This will affect the numerical values for $B^q(0)$, but 
the total anomalous gravitomagnetic moment of the proton is always zero, and 
the Ji sum rule is always correctly satisfied. 

For completeness, we remark that orbital angular momentum can also be accessed 
through the twist-3 GPD $G_2^q(x,\xi,t)$ \cite{Penttinen:2000dg,Kiptily:2002nx}. 
The connection of this GPD to orbital angular momentum in the bag model was studied
in \cite{Courtoy:2016des}. Another expression for the orbital angular momentum
involving two twist-3 GPDs, $G_2^q(x,\xi,t)$ and $G_4^q(x,\xi,t)$, was proposed in 
Ref.~\cite{Hagler:2003jw}.

\newpage
\section{Orbital angular momentum and the pretzelosity GTMD}
\label{Sec-8}

The pretzelosity TMD $h_{1T}^{\perp q}(x,k_T^2)$ is related in certain quark models
to orbital angular momentum. It is instructive to briefly review the model 
studies which lead to this insight. In the light-cone representation and taking 
into account effects of Melosh-Wigner rotations, it was observed in
Ref.~\cite{Ma:1998ar} that the difference between the transversity PDF $h_1^q(x)$ 
and helicity PDF $g_1^q(x)$ is related to quark orbital angular momentum. 
In Ref.~\cite{Avakian:2008dz} it was shown that, in the bag model, 
the difference of helicity and transversity TMDs is given in terms 
of the transverse moment of pretzelosity TMD, 
$g_1^q(x,k_T^2)-h_1^q(x,k_T^2)=h_{1T}^{\perp(1)q}(x,k_T^2)$ where
$h_{1T}^{\perp(1)q}(x,k_T^2)=\frac{k_T^2}{2M^2}h_{1T}^{\perp q}(x,k_T^2)$.
This in turn paved the way to the insight that pretzelosity is related
to quark orbital angular momentum as \cite{She:2009jq}
\begin{equation}\label{eq:pretzel-1}
    \LJM_z^q=-\int dx\int d ^2\vec{k}_T \;\frac{\vec{k}_T^{\,2}}{2M^2}\,
    h_{1T}^{\perp q}(x,\vec{k}^2_T)\, . 
\end{equation}
This result is supported in the 
light-cone SU(6) quark-spectator model \cite{She:2009jq},
bag model \cite{Avakian:2010br},
covariant parton model \cite{Avakian:2010nz,Efremov:2010cy},
light-front constituent quark model \cite{Lorce:2011kd}, and
chiral quark-soliton model restricted to three-quark sector~\cite{Lorce:2011kd}.
Models like the spectator model of Ref.~\cite{Liu:2014zla} or the light-front 
models of Refs.~\cite{Chakrabarti:2016yuw} do not 
support Eq.~(\ref{eq:pretzel-1}). Spherical symmetry of quark wave functions 
was shown to be an important prerequisite for Eq.~(\ref{eq:pretzel-1}) to hold 
in a quark model \cite{Lorce:2011kn}. In models where this symmetry is not 
present and in QCD, Eq.~(\ref{eq:pretzel-1}) is not supported \cite{Boussarie:2023izj}. 

Although not general, it is nevertheless interesting to ask the question why 
the Eq.~(\ref{eq:pretzel-1}) holds in certain models. The observation that 
some model symmetry must be present \cite{Lorce:2011kn} sheds light on a necessary 
technical requirement. But it would be interesting to gain further insights. For that
we inspect the $x$-dependence. In the bag model, the expression for the pretzelosity TMD 
(rewritten in the notation of this work) is given by 
\cite{Avakian:2010br}
\be\label{Eq:pretzel-def}
    h_{1T}^{\perp{q}}(x,\vec{k}_T^{\,2}) = P^q A^2 M^3 \bigl[ \, -2t_1^2\, \bigr]. 
\ee
Comparing to the model expression of the GTMD 
$F_{1,4}^q(x, \, \xi, \, \vec{k}_T^{\,2}, \vec{k}_T\cdot\vec{\Delta}_T,\, \vec{\Delta}_T^{\,2})$ 
in Eq.~(\ref{Eq:F14-model}) in the limit $\xi\to0$ and $\vec{\Delta}_T\to0$, we find in the 
bag model the following relation
\be\label{eq:pretzel-2}
    2 F_{1,4}^q(x,0,\vec{k}_T^{\,2},0,0) = h_{1T}^{\perp q}(x,\vec{k}_T^{\,2}).
\ee
This is an interesting observation. It shows that the relation is deeper than what one
might conclude from the mere observation that the same model expression $\LJM_z^q$ results 
from integrating $2 F_{1,4}^q(x,0,\vec{k}_T^{\,2},0,0)$ and $h_{1T}^\perp(x,\vec{k}_T^{\,2})$
over $dx$ and $d^2k_T$ in Eqs.~(\ref{OAM},~\ref{eq:pretzel-1}). In fact, the relation holds 
at the level of the integrands in Eqs.~(\ref{OAM},~\ref{eq:pretzel-1}) before integrations 
over $x$ and transverse parton momenta are carried out. 

At this point, one may wonder whether the relation of pretzelosity and orbital angular momentum
in the bag model could be still deeper. In fact, the pretzelosity TMD can be obtained as the 
forward limit of one of the GTMDs related to the Dirac structure $\Gamma = i\sigma^{j+}\gamma_5$. 
This Dirac structure defines 8 leading GTMDs $H_{1,l}^q$ with $l=1,\;\dots\,, \; 8$. The GTMD 
containing the pretzelosity TMD as a forward limit is $H_{1,4}^q$, and the relation is given by
\be\label{Eq:pretzel-GTMD-relation}
    h_{1T}^{\perp{q}}(x,\vec{k}_T^2) = H_{1,4}^q(x,0,\vec{k}_T^2,0,0)\,.
\ee
In order to investigate if there is deeper relation, it is instructive 
to derive the bag model expression for $H_{1,4}^q$ by repeating the steps 
in Eqs.~(\ref{Eq:correlator-GTMDs-large-Nc}-\ref{Eqs:F1j-model}) for 
$\Gamma = i\sigma^{j+}\gamma_5$. Details of this calculation 
(and results for the other $H_{1,l}^q$, $l\neq 4$)
will be reported elsewhere \cite{new}. Here we shall content ourselves
with quoting the result for $H_{1,4}^q$ which is given by
\be \label{eq:H14-model}
    H_{1,4}^q = P^q A^2 M^3 \Bigl[-2\,t_1^+ \, t_1^- \Bigr].
\ee
{In the forward limit, we reproduce Eq.~(\ref{Eq:pretzel-GTMD-relation}) 
from (\ref{eq:H14-model}).}
A comparison of Eqs.~(\ref{Eq:F14-model}) and (\ref{eq:H14-model}) 
yields the relation
\be \label{eq:H14andF14}
 H_{1,4}^q(x, \, \xi, \, \vec{k}_T^{\,2}, \vec{k}_T\cdot\vec{\Delta}_T,\, \vec{\Delta}_T^{\,2})
 =2F_{1,4}^q(x, \, \xi, \, \vec{k}_T^{\,2}, \vec{k}_T\cdot\vec{\Delta}_T,\, \vec{\Delta}_T^{\,2}).
\ee
This is the desired result revealing a deeper connection: in the bag model,  
pretzelosity appears to be connected to orbital angular momentum because the 
GTMDs $F_{1,4}^q$ and $H_{1,4}^q,$ as functions of all $5$ variables, are related to 
each other according to Eq.~(\ref{eq:H14andF14}). To the best of our knowledge, 
this relation has not been observed in a model before.

It will be interesting to explore along the lines of Ref.~\cite{Lorce:2011kn}
whether some symmetry of quark wave functions might be a necessary (sufficient) 
condition for the existence of the relation (\ref{eq:H14andF14}) in a model, 
and whether there are potentially other models which support such a relation.

\newpage
\section{Conclusions}
\label{Sec-9}

We presented a study of the leading GTMDs $F^q_{1,1}$, $F^q_{1,2}$, $F^q_{1,3}$, $F^q_{1,4}$
associated with $\Gamma=\gamma^+$ in the bag model in the large-$N_c$ limit. 
The results satisfy Hermiticity properties and correctly reproduce known relations to TMDs and GPDs 
with $F^q_{1,1}$ relating to the TMD $f_1^q(x,\vec{k}_T^2)$ and $F^q_{1,1}$, $F^q_{1,2}$, $F^q_{1,3}$ 
relating to the GPDs $H^q(x,\xi,t)$ and $E^q(x,\xi,t)$. This demonstrates, within 
the limitations of the model, the theoretical consistency of the bag model. 
This includes the polynomiality property, which was not explicitly discussed in 
this work, but verified in \cite{Ji:1997gm,Tezgin:2024tfh}.

Only $F^q_{1,4}$ was studied previously in the bag model \cite{Courtoy:2016des}. 
Our results for $F^q_{1,4}$ are in agreement with those of~Ref.~\cite{Courtoy:2016des}.
To the best of our knowledge, the GTMDs $F^q_{1,1}$, $F^q_{1,2}$, $F^q_{1,3}$ have not been 
studied in the bag model before and constitute novel results. We applied the obtained results 
to study the orbital angular momentum distributions as defined via $F^q_{1,4}$ and the 
Ji sum rule. As observed in prior model studies, the $x$-dependencies of these two partonic
distributions differ, but the integrated contributions $\LJM_z^q$ coincide, which is expected 
in quark models. The numerical results for the $u$- and $d$-quark contribution to $\LJM_z^q$
agree with what is typically found in quark models at low hadronic~scales. We furthermore 
obtained several interesting new insights.

One interesting new insight is based on our analytical proofs for the total quark angular momentum 
sum rule for the $J^q$ contribution to the nucleon spin 
(i) as the sum of quark orbital angular momentum (obtained via $F_{1,4}^q$) and quark spin 
(obtained via the helicity PDF $g_1^q$) which is often referred to as the Jaffe-Manohar spin sum rule, and 
(ii) using the Ji spin sum rule.
Both spin sum rules were previously tested numerically in the bag model within an accuracy of better than 
$1\,\%$, the former in Ref.~\cite{Courtoy:2016des} and the latter in Refs.~\cite{Ji:1997gm,Neubelt:2019sou,Courtoy:2016des}.

Analytical proofs show something numerical tests cannot reveal. 
The Jaffe-Manohar spin sum rule is satisfied in the bag model based on quark wave function
normalization and spin-flavor assignment in SU(4) spin-flavor symmetry and is analogous to
the quark flavor sum~rule, which can be proven in a similar way. 
The proof of the Ji sum rule is more sophisticated, 
and requires {the explicit use of the equation of motion, 
i.e.\ it invokes the dynamics in the model.} From this point of view, the proof of the Ji spin 
sum rule resembles that of the momentum sum rule. It would be interesting to see whether similar 
distinctions between Jaffe-Manohar and Ji sum rules can also be made in other models.\footnote{After
    our work was completed, we learned that the situation is similar (but not quite the same) in 
    the scalar diquark model of~\cite{Lorce:2017wkb}.} 

Another interesting insight concerns the connection between orbital angular momentum and the 
pretzelosity TMD. It has been observed in several quark models that the $x$-integral of the 
(1)-transverse moment of the pretzelosity TMD is related to $\LJM_z^q$. In this work, we have 
shown that this relation is deeper and holds on the level of the $x$-integrands. In fact, 
in the bag model, the pretzelosity TMD $h_{1T}^{\perp q}(x,\vec{k}{ }^2_T)$ coincides with 
$2F_{1,4}^q(x,0,\vec{k}{ }^2_T,0,0)$ in the forward limit. Finally, we went one step further 
and demonstrated that, in the bag model, the relation $H_{1,4}^q=2F_{1,4}^q$ holds even when 
all the GTMD variables $x$, $\xi$, $\vec{k}{ }^2_T$, $\vec{k}{ }_T\cdot\vec{k}{ }^2_T$ and 
$\vec{\Delta}{ }^2_T$ are respectively kept non-zero or are not integrated over. 
Here, $H_{1,4}^q$ is the GTMD which contains the pretzelosity TMD as a limiting case in the 
forward limit. To the best of our knowledge, this extension of the relation between pretzelosity 
and orbital angular momentum in terms of GTMDs has not been encountered in model studies before. 

It is important to keep in mind that models have limitations. Such relations can be found unsupported 
in quark models exhibiting a lesser degree of symmetry or can be spoiled by relaxing certain model
assumptions. Models with gauge field degrees of freedom cannot be expected to support such relations, 
and in QCD all GTMDs are independent functions and describe complementary aspects of nucleon structure. 
However, quark models have been found to catch some gross features of TMDs and GPDs observed in 
phenomenology or lattice QCD studies. It would be therefore interesting to see whether this might be 
true also in the case of the GTMDs $F_{1,4}^q$ and $H_{1,4}^q$ --- at least in some region of the variables,
for instance, valence-like $x$, small $\vec{k}{ }_T^2 < M^2$ or moderate $\vec{\Delta}{ }_T^2 < M^2$. 

It will be therefore interesting to see if our findings can be confirmed in other quark model studies
and perhaps even in lattice QCD --- exactly or as approximate relations. The ultimate testing ground 
will be provided by phenomenological studies. The path to rigorous factorization proofs and phenomenological 
tests of GTMDs is long. Until then such model relations can be {helpful guidelines} to prepare the first 
estimates for GTMD-related observables. We can look forward to learn more about non-perturbative properties
of GTMDs from models.

\ \\
{\bf Acknowledgments.} The authors wish to thank Shohini Bhattacharya, C\'edric Lorc\'e, 
Barbara Pasquini, and Emaan Sohail for valuable discussions. This work was supported by 
NSF under the Award No. 2412625, and DOE under the umbrella of the Quark-Gluon Tomography 
(QGT) Topical Collaboration with Award No. DE-SC0023646.

\newpage
%%%%%%%%%%%%%%%%%%%%%%%%%%%%%%%%%%%%%%%%%%%%%%%%%
\appendix
%%%%%%%%%%%%%%%%%%%%%%%%%%%%%%%%%%%%%%%%%%%%%%%%%

% \section{Proof of \eqref{Eq:DerivativeofF14} being equivalent to $\widetilde{E}_{2 T}+H+E$}
% \label{App:DerivativeofF14Proof}
% Combing \eqref{Eq:tildeE2T} with \eqref{Eq:F27-model} and \eqref{Eq:F28-model} and taking the forward limit of  $\widetilde{E}_{2 T}$ gives
% \be
%    \widetilde{E}_{2 T}^q(x,0,0)=-2P^qA^2M^2\int d^2 \vec{k}_T \biggl[
%     \biggl(\frac{\vec{k}_T^2}{2k}(t_0t_1'-t_0't_1)+t_0t_1\biggr)\biggr]
% \ee
% from which, combing with \eqref{Eq:GPD-forward-limit} and \eqref{Eq:GPD-E-forward-limit}, gives the result
% \be
%  \widetilde{E}_{2 T}^q(x,0,0)+H^q(x,0,0)+E^q(x,0,0)=2P^qA^2M^2\int d^2 \vec{k}_T (M x-\epsilon_0)t_1^2
% \ee

%====== REFERENCES =========================================================

\end{document}